# Photonic Floquet skin-topological effect

Yeyang Sun[1#], Xiangrui Hou[1#], Tuo Wan[1], Fangyu Wang[1], Shiyao Zhu[1,2], Zhichao Ruan[1*] and Zhaoju Yang[1*]

[1]School of Physics, Interdisciplinary Center for Quantum Information, Zhejiang Province Key Laboratory of Quantum Technology and Device, Zhejiang University, Hangzhou 310027, Zhejiang Province, China

[2]Hefei National Laboratory, Hefei 230088, China

#These authors contributed equally to this work

*Email: zhichao@zju.edu.cn; zhaojuyang@zju.edu.cn.

**Abstract**

Non-Hermitian skin effect and photonic topological edge states are of great interest in non-Hermitian physics and optics. However, the interplay between them is largly unexplored. Here, we propose and demonstrate experimentally the non-Hermitian skin effect that constructed from the nonreciprocal flow of Floquet topological edge states, which can be dubbed 'Floquet skin-topological effect'. We first show the non-Hermitian skin effect can be induced by pure loss when the one-dimensional (1D) system is periodically driven. Next, based on a two-dimensional (2D) Floquet topological photonic lattice with structured loss, we investigate the interaction between the non-Hermiticity and the topological edge states. We observe that all the one-way edge states are imposed onto specific corners, featuring both the non-Hermitian skin effect and topological edge states. Furthermore, a topological switch for the skin-topological effect is presented by utilizing the gap-closing mechanism. Our experiment paves the way of realizing non-Hermitian topological effects in nonlinear and quantum regimes.

Topological insulators are a new phase of matter that is constituted by insulating bulk and conducting edges. They have been extensively explored in condensed matter physics [1,2], photonics [3–11], phononics [12–14], and so on. In photonics, shortly after the observation of topologically protected edge states in microwaves [4], the topological states in the optical frequency range relying on artificial gauge fields were experimentally realized [5,6]. One paradigmatic example is the photonic Floquet topological insulator [5] consisting of a honeycomb array of helical optical waveguides. The periodic driving force results in the artificial gauge field and Floquet topological phases. The one-way topological edge states that are immune to backscattering were predicted and observed. The realizations of the topological states in classical-wave systems show potential in lasing [15–18] and quantum sources [19,20].

Characterized by complex eigenenergies and nonorthogonal eigenstates, non-Hermitian physics [21–23] governing systems interacting with the environment has led to many frontiers, such as PT symmetric physics [24–34] and non-Hermitian topological phases [35–44]. Recently, the non-Hermitian skin effect (NHSE) [45–58] has drawn a lot of attention both in theory and experiment. The NHSE features the coalescence of the extended bulk states into the edges of 1D systems, which can be well described by the non-Bloch band theory [48,51]. The interplay between the NHSE and aforementioned topological states brings us a new concept of the hybrid skin-topological effect [59–61]. Different from the higher-order non-Hermitian skin effect [62–66], the action of the coalescence of extended eigenstates for the skin-topological effect is only on the topological edge modes [60,61]. Therefore, the number of the skin-topological modes is proportional to the length size of the system. So far, such an effect has only been realized by introducing asymmetric coupling into higher-order topological insulators [66], whereas the interaction between the one-way propagating topological edge states and NHSE remains less explored.

In this work, we bridge this gap by adopting a 2D optical array of lossy helical waveguides and observe the photonic Floquet skin-topological effect. First, by introducing staggered loss into a 1D optical array of helical waveguides, we realize the Floquet non-Hermitian skin effect. Next, we pile up the 1D lattice and arrive at a 2D non-Hermitian Floquet topological insulator. The complex spectrum of the non-Hermitian system shows that the gapless unidirectional edge states spanning across the topological band gap (corresponding to a nonzero Chern number of 1) can acquire the non-trivial point-gap winding topology [45,60,61,67], which indicates the existence of the NHSE that induced by the nonreciprocal flow of these edge states. The sign of the winding number determines which corner of the sample is the topological funnel of light. Moreover, by introducing a large enough on-site energy difference and closing the Floquet topological band gap, a topological switch [61,68] for the Floquet skin-topological effect is demonstrated.

We start from a 1D optical array consisting of helical waveguides, as shown in Fig. 1a. This 1D optical array contains two sublattices A and B, and the sublattice B is endowed with considerable loss, as marked in blue. The paraxial propagation of light in this non-Hermitian system can be described by the tight-binding equation [5]:

$$i\partial_z \psi_n = \sum_{\langle m \rangle} c e^{iA(z)\cdot r_{mn}} \psi_m - \sum_{j \in A} i\gamma \psi_j = H(z)\psi_n \quad (1)$$

where $\psi_n$ is the electric field amplitude in the *n*th waveguide, $c$ is the coupling strength between the two nearest waveguides, $r_{mn}$ is the displacement pointing from waveguide *m* to *n*. This *z*-dependent equation describing paraxial light propagation can be mapped to a time-dependent Schrodinger equation and the *z* axis plays the role of time. The periodic driving is equivalent to adding a time-dependent vector potential $A(z) = kR\Omega(sin(\Omega z), -cos(\Omega z), 0)$ to the optical array. The distance between the two nearest waveguides is $a = 15$ μm and the lattice constant is $d = 15\sqrt{3}$ μm. The helix radius is $R = 8$ μm and the period is $Z = 1\ cm$.

For the above tight-binding model with *z*-periodic Hamiltonian, its eigenstates can be calculated from the equation of $H_{eff}\psi = \beta\psi$, where $H_{eff} = \frac{1}{Z}\int_0^Z H(z)dz$ is the effective Hamiltonian for wavefunctions over one period and $\beta$ is the quasienergy for Floquet systems. The complex quasienergy spectrum of this 1D optical array under periodic boundary condition (PBC) and open boundary condition (OBC) is shown in Fig. 1b, as labeled by grey dotted curves and black dots, respectively. We can see that the complex spectrum under PBC forms a closed loop and drastically collapses into a line under OBC. The eigenfunctions displayed in Fig. 1b reveal that the eigenstates all localize at the left boundary, which is the direct result of the NHSE. The closed loop in the complex spectrum results in the non-trivial point-gap topology, which can be characterized by the winding number [45,69] for Floquet systems

$$w = \int_0^{2\pi} \frac{dk}{2\pi} \partial_k \arg[\beta(k) - \beta_0] \qquad (2)$$

where $\beta_0$ is a reference quasienergy for numerical calculations. The winding number for the 1D optical array is $w = 1$ indicating the existence of the NHSE.

In experiments, we fabricate the 1D optical array of helical waveguides by utilizing the femtosecond laser writing method [5]. The optical loss in sublattice B is introduced by setting breaks periodically into the waveguides [70] (see methods for more details). To observe the 1D Floquet NHSE, a laser beam with a wavelength of 635 nm is initially launched into the center waveguide of the 1D array. We perform a series of measurements at different propagation lengths of *z*=2, 4, 6, 8 and 10 cm. The results are shown in Fig. 1d. The white dashed circles mark the location of the input waveguide. As we can see, the light propagates continuously to the left, which indicates the collapse of the extended eigenstates onto the left boundary. The overall reduced light intensity as increasing the propagation length is due to the passive setting of the optical array. This observation unravels the existence of 1D NHSE induced by pure loss as well as periodic driving and provides a cornerstone for the next exploration of the interplay between the NHSE and photonic topological edge states.

Having observed the Floquet NHSE in a 1D array, we investigate the interaction between the non-Hermiticity and topological edge states. We pile up the 1D Floquet lattice composed of helical waveguides and arrive at a 2D photonic non-Hermitian Floquet topological insulator with the nonzero non-Hermitian Chern number [35,67] of $C = 1$. The structured loss is introduced into one sublattice of the honeycomb lattice. The non-trivial topology guarantees the existence of the topological boundary states when we consider a finite sample. The complex spectra for two cases of *x*-PBC/*y*-OBC (periodic boundary along the *x*-axis and open boundary in the *y*-direction, as pointed out in panel d) and *x*-OBC/*y*-OBC are shown in Fig. 2a,c. Apart from the bulk states marked by grey dotted points, there exist two counter-propagating edge states localized at the opposite edges spanning across the topological band gap in the complex spectrum, as displayed in Fig. 2a, b. The blue (red) dots correspond to the right (left)-propagating edge mode with relatively larger (negligible) loss. The complex spectral loop in the topological band gap reveals the non-trivial point-gap topology characterized by the winding number of $w = 1$, which results in the NHSE induced by the one-way edge flow of light. For the case of double OBC (x-OBC/y-OBC), the closed loop in the complex spectrum drastically collapses (Fig. 2c) and all non-Hermitian topological edge modes localize at the upper-left corner (labeled as corner I in Fig. 2d), whereas the bulk states stay extended. The solid (dashed) white arrows pointing to the propagating direction of the edge states correspond to the flow of light with negligible (large) loss. As a result, all the energy of the topological edge states accumulates at corner I, which indicates the existence of the so-called hybrid skin-topolgoical modes. Note that if we add the loss configuration on the sublattice B, the point-gap winding will change to $w = -1$, which gives rise to the skin-topological modes at corner III.

To experimentally study the Floquet skin-topological effect, a 2D non-Hermitian honeycomb lattice of helical waveguides is fabricated. The breaks that can generate relatively larger optical loss are introduced into the waveguides of sublattice A. To excite the topological edge modes, a broad tilted Gaussian beam with the momentum of $kd = \pi$ is initially launched into the outer perimeter of the sample. The white ellipse indicates the position and shape of the launched beam. The total propagation length in the sample is 10 cm (z-axis). By moving the injected beam along the outer perimeter, we can see in Fig. 3a,b that the input light can propagate unidirectionally along the edge of IV-III, circumvent corner III (panel a), and propagate along the edge of III-II, circumvent the sharp corner II (panel b) without backscattering. However, the situation changes drastically when the input light encounters corner I. By moving the input Gaussian beam leftwards along the edge, as shown in Fig. 3c, we observe that the injected light propagates along the upper edge and stays trapped at corner I without penetration into the bulk, which is the direct evidence of the existence of the hybrid Floquet skin-topological modes. The observations in Fig. 3 together elucidate that the

Floquet skin-topological effect features both the NHSE and topological protection of the propagating light.

For comparison, we fabricate a 2D Hermitian photonic Floquet topological insulator with no structured loss. We reproduce the similar experiments and observe that the initial tilted wave packet can propagate counterclockwise along the outer perimeter and bypasses corner I (see methods for more details).

A topological switch [59,68] manifests that topology can provide a switch for the NHSE through topological phase transitions. In our model, by introducing a large enough on-site energy difference between the two sublattices (see methods for more details), the topological phase transition occurs and the Flouqet topological band gap closes. Without the topological edge states providing asymmetric coupling for the NHSE, the complex spectrum shows no point-gap winding and therefore no Floquet skin-topological modes can be found. We adopt the same experimental philosophy as shown in Fig. 3. As we can see in Fig. 4a-c, the injected light penetrates into bulk and cannot accumulate at corner I. Therefore, the skin-topological effect is switched off.

In conclusion, we have experimentally demonstrated the Floquet NHSE in a 1D lossy optical array and Floquet skin-topological effect in a 2D non-Hermitian photonic Floquet topological insulator. By introducing structured loss into the periodically driven optical waveguides, we have found the point-gap topology in the complex spectra of the 1D and 2D non-Hermitian photonic systems. In experiments, we have fabricated the optical lattices by the standard femtosecond laser-writing method and observed the topological funnel of light at the left boundary and corner I of the 1D and 2D optical arrays, respectively. Moreover, a key to switching on/off the skin-topological effect has been realized by utilizing the topological phase transition. Our work investigates the interaction between the NHSE and photonic topological edge states and provides the first example of the NHSE in an optical Floquet topological insulator, which may pave the way for further exploration of non-Hermitian topological effects [71,72] in nonlinear [73–75] and quantum regimes [76,77].

**Acknowledgements**

This research is supported by the National Key R&D Program of China (Grant No. 2022YFA1404203, 2022YFA1405200), National Natural Science Foundation of China (Grant No. 12174339, 12174340), Zhejiang Provincial Natural Science Foundation of China (Grant No. LR23A040003) and Excellent Youth Science Foundation Project (Overseas).

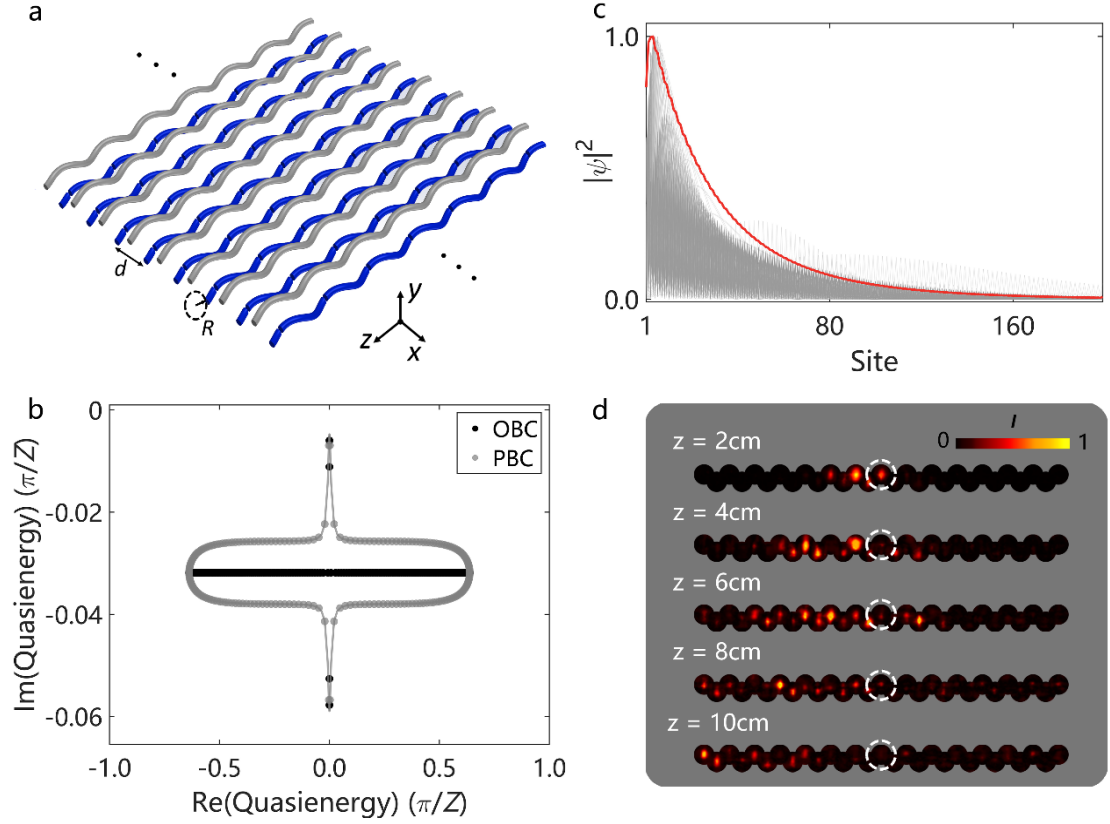

Figure 1. Floquet NHSE in a 1D optical array. a, Schematic of a 1D non-Hermitian optical array consisting of helical waveguides. The color grey (blue) represents normal (lossy) waveguides. The optical loss is introduced by setting breaks, as can be seen in the blue waveguides. b, Complex quasi-energy spectrum of the 1D optical array. The spectral loop under PBC indicates the existence of the point-gap topology. c, Floquet skin eigenmodes (grey curves) and the summing eigenmodes (red curve). We set 200 sites in the array for these numerical results. The parameters for simulations are: coupling strength $c = 1.5 \text{ cm}^{-1}$, optical loss $\gamma = 0.2 \text{ cm}^{-1}$, lattice constant $d = 15\sqrt{3} \text{ μm}$, helix radius $R = 8 \text{ μm}$ and the period $Z = 1 \text{ cm}$. d, Experimental observation of the Floquet NHSE. The light shifts continuously to the left indicating the collapse of the eigenstates into the left boundary.

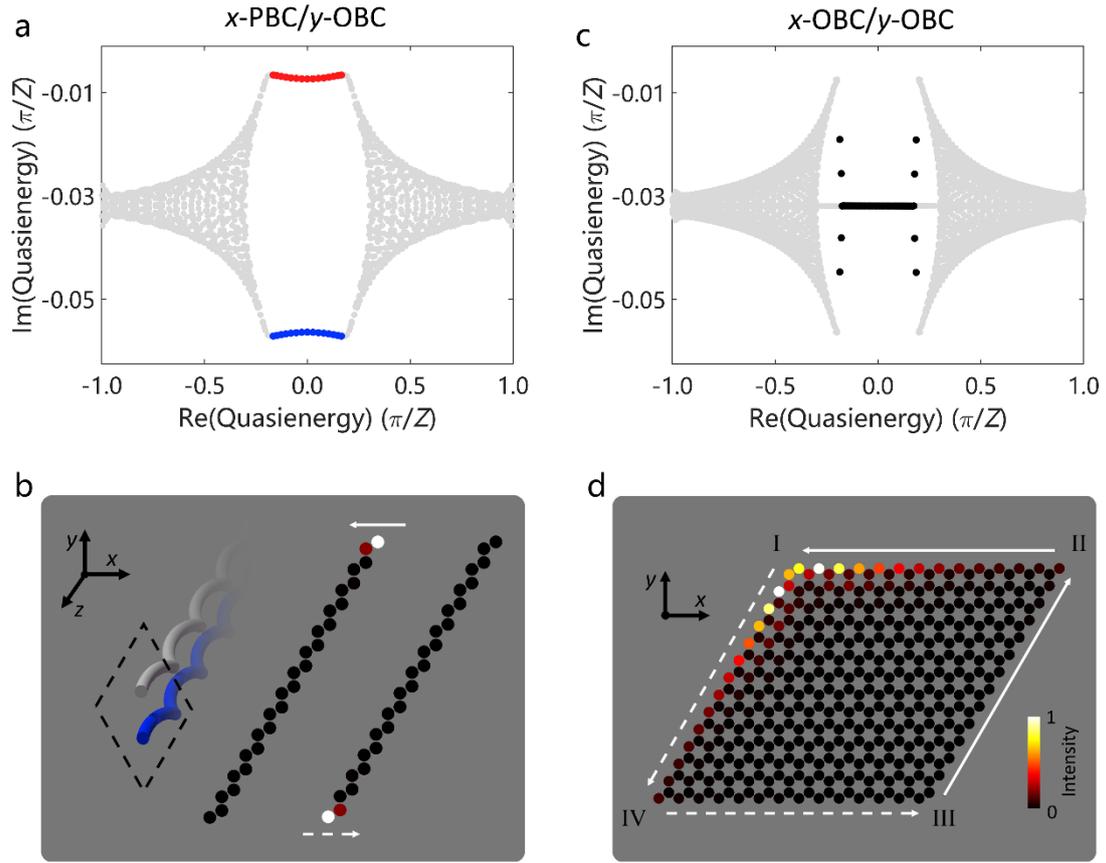

Figure 2. Hybrid skin-topological effect in a non-Hermitian 2D photonic Floquet topological insulator. a, b, Complex quasienergy spectrum under x-PBC/y-OBC. The blue (red) dots correspond to the right (left)-propagating edge mode with relatively larger (negligible) loss. The spectral loop in the topological band gap reveals the non-trivial point-gap topology indicating the existence of the NHSE of light at the boundary. c, d, Complex quasienergy spectrum under *x*-OBC/*y*-OBC. The skin-topological modes (black dots) emerge and localize at the upper-left corner of the sample (corner I).

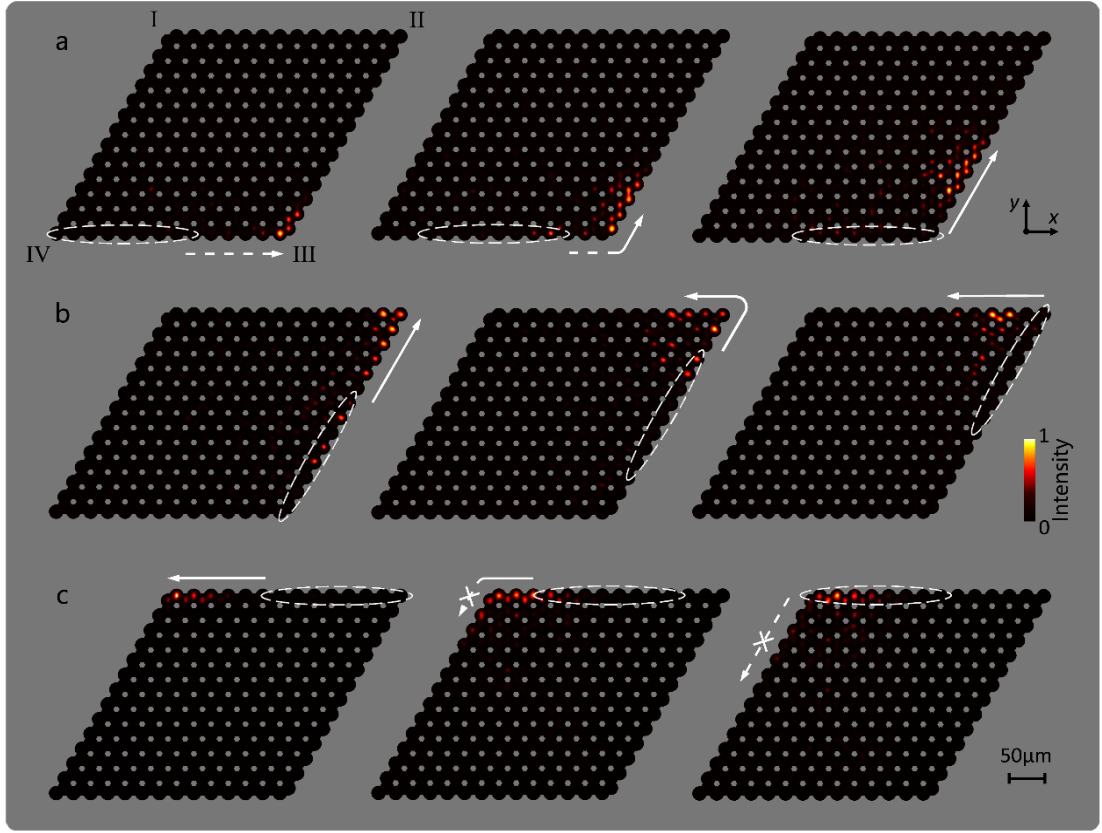

Figure 3. Experimental observation of the Floquet skin-topological effect. By moving the input tilted Gaussian beam along the outer perimeter, we observe the output light distribution at the end facet of the sample after 10 cm long propagation. a, The injected light propagates along the edge of IV-III and bypasses corner III without backscattering. b, The light propagates along the edge of III-II and bypasses the sharp corner III without backscattering. c, The light propagates along the edge of II-I and accumulates at corner I, which confirms the existence of the Floquet skin-topological modes.

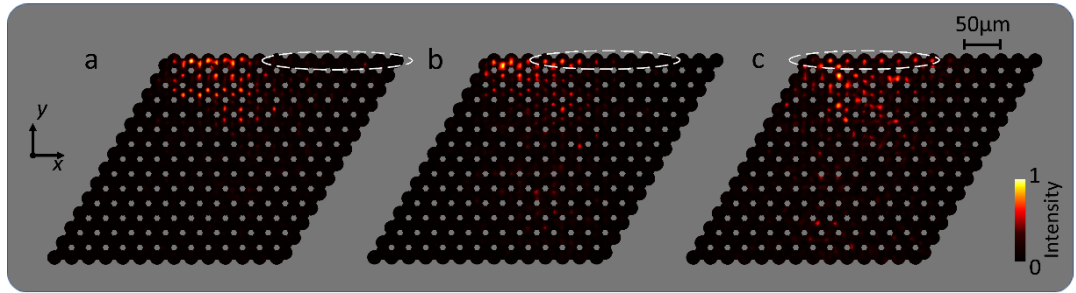

Figure 4. Topological switch for the skin-topological effect. a-c, Introducing large enough on-site energy difference of $\Delta m = 3c$ results in the topological phase transition and closes the Flouqet topological band gap. In experiments, by moving the input tilted Gaussian beam leftwards along the upper edge, we can see that the injected light penetrates into the bulk. The skin-topological effect is switched off by the gap-closing mechanism.